\documentclass[twocolumn,prl]{revtex4}
\newcommand\p{\partial}
\newcommand\CD{{\cal D}}
\newcommand\CT{{\cal T}}
\newcommand\CV{{\cal V}}
\newcommand\be{\begin{equation}}
\newcommand\ee{\end{equation}}
\newcommand\bea{\begin{eqnarray}}
\newcommand\eea{\end{eqnarray}}
\renewcommand\tilde[1]{\widetilde{#1}}
\renewcommand\hat[1]{\widehat{#1}}
\begin{document}
%%%%%%%%%%%%%%%%%%%%%%%%%%%%%%%%%%%%%%%%%%%%%%%%%%%%%%%%%%%%%%%%%%%%%%%%%%%%%%
\title{Closed-String Tachyon Condensation and the Worldsheet
Super-Higgs Effect}
\author{Petr Ho\v rava and Cynthia A. Keeler}
\affiliation{Berkeley Center for Theoretical Physics and Department of
Physics,\\
University of California, Berkeley, California 94720-7300\\
and\\
Physics Division, Lawrence Berkeley National Laboratory,\\
Berkeley, California 94720-8162, USA}
\begin{abstract}
Alternative gauge choices for worldsheet supersymmetry can elucidate dynamical 
phenomena obscured in the usual superconformal gauge.  In the particular 
example of the tachyonic $E_8$ heterotic string, we use a judicious gauge 
choice to show that the process of closed-string tachyon condensation can be 
understood in terms of a worldsheet super-Higgs effect.  The worldsheet 
gravitino assimilates the goldstino and becomes a dynamical propagating 
field.  Conformal, but not superconformal, invariance is maintained throughout.
\end{abstract}
\pacs{}
\maketitle
%%%%%%%%%%%%%%%%%%%%%%%%%%%%%%%%%%%%%%%%%%%%%%%%%%%%%%%%%%%%%%%%%%%%%%%%%%%%%%
In theories with local worldsheet supersymmetry, it is traditional to work in 
superconformal gauge, sometimes supplemented by a subsidiary gauge-fixing 
condition such as light-cone gauge.  This superconformal gauge choice has been 
successful in a large class of static backgrounds, but it may obscure aspects 
of more complicated systems. 

String theory has now reached the stage which requires answers to a new class 
of dynamical questions, involving backgrounds far from equilibrium and with 
substantial time dependence.   Such questions are particularly pressing 
in the cosmological setting.  This challenge is likely to require new 
techniques.  

We propose using a meaningful alternative to superconformal gauge; this
previously unexplored tool can lead to new insights into string dynamics.
To illustrate this phenomenon, we focus on a particular 
example which epitomizes time-dependent processes:  The problem of 
closed-string tachyon condensation.  While understanding the dynamics of its 
open-string cousins has improved our picture of brane decay into the vacuum 
(or to lower-dimensional stable branes), the closed-string tachyon is related 
to the possible decay of spacetime itself.  This makes the problem difficult, 
and highlights the current 
limitations in our undestanding of string theory far from equilibrium.

In this letter, we demonstrate the use of alternative worldsheet gauge choices
in the specific example of a ten-dimensional heterotic string model with $E_8$
gauge symmetry and a singlet tachyon.  This little-studied model belongs to the
early classification of modular invariant string theories in ten
dimensions \cite{klt}, but its {\it raison d'\^etre\/} has been hitherto
obscure.  In a more detailed companion paper \cite{pixie}, we argue that
this $E_8$ heterotic string describes the fate of an unstable
configuration in heterotic M-theory with the two $E_8$ boundaries
breaking complementary sets of sixteen supercharges \cite{fh}.  This
configuration suffers from the ``decay to nothing'' instability
(at least in supergravity), due to an instanton connecting the two
boundaries by a throat.  Since there is an attractive Casimir force
between the two $E_8$ boundaries, the fate of the ``decay-to-nothing''
instability should be addressed at weak string coupling where it turns
into a perturbative tachyonic instability.  The unique tachyonic $E_8$
string is the only promising candidate for describing this decay.

In the present work we put this motivation aside, and study the
tachyon decay in the weakly coupled heterotic $E_8$ model as an
independently interesting problem.
The worldsheet theory of the heterotic $E_8$ string consists of $(0,1)$ chiral
supergravity, coupled to matter supermultiplets.  In the fermionic
representation, the worldsheet action of the $E_8$ string is locally identical
to that of its supersymmetric cousins,
\bea
S_0&=&\frac{1}{4\pi\alpha'}\int d^2\sigma e\left\{(h^{mn}\p_m X^\mu\p_n X^\nu
+\frac{i}{2}\psi^\mu\gamma^mD_m\psi^\nu\right.\nonumber\\
&&{}+i\chi_m\gamma^n\gamma^m\p_nX^\mu\psi^\nu)\eta_{\mu\nu}
+\frac{i}{2}\lambda^A\gamma^mD_m\lambda^A\nonumber\\
&&\left.\vphantom{\frac{1}{2}}\qquad\qquad{}+F^AF^A\right\}.
\eea
The matter fermions are chiral, with $\Gamma\lambda=\lambda$,
$\Gamma\psi=-\psi$ where $\Gamma\equiv\gamma^0\gamma^1$, and the gravitino
satisfies $\Gamma\chi_m=\chi_m$.  We are in Minkowski signature, with
$h_{mn}=e_m{}^ae_n{}^b\eta_{ab}$.

This tachyonic $E_8$ model is distinguished by its GSO projection:
It is the only one in which no two fermions among $\lambda^A$ carry
always the same spin structure.  It turns out that 31 of the
$\lambda^A$s give rise to the level-two $E_8$ current algebra, of
left-moving central charge $c_L=31/2$. We denote by $A'$ the index
that runs over those first 31 values of $A$. The remaining $1/2$ of
central charge is supplied by the remaining fermion
$\lambda^{32}\equiv\lambda$.  The spin structure of $\lambda$ is
always the same as that of the right-moving NSR fermions $\psi^\mu$,
a property that singles out $\lambda$ uniquely. As we shall see, this
``lone fermion'' $\lambda$ plays a central role in the process of
tachyon condensation.

We wish to examine the spacetime evolution of the tachyon condensate
as an on-shell process in string theory.  Thus our worldsheet theory
must maintain conformal invariance.  Accordingly, we turn on a
linear dilaton background $\Phi=V_\mu X^\mu$. Since we do not wish
to change the critical dimension, the dilaton gradient $V$ is chosen
to be null. We choose spacetime lightcone coordinates $X^\pm=X^0\pm
X^1$ (with the spacetime index $\mu$ splitting as $(\pm,i)$, $i=1,\ldots 8$),
such that the only nonzero component of $V$ is $V_-$.  The action is
$S_0+S_V$, where
\be
S_V=-\frac{1}{4\pi}\int d^2\sigma e\,V_-\left(X^-R+2i\chi_m\gamma^n
\gamma^mD_n\psi^-\right).
\ee
This is invariant under local supersymmetry transformations, given by
\bea
\delta e_m{}^a&=&2i\epsilon\gamma^a\chi_m,\quad
\delta\chi_m=D_m\epsilon\nonumber\\
\delta X^\mu&=&i\epsilon\psi^\mu,\quad\delta\psi^\mu=\gamma^m\p_m X^\mu\epsilon
+\alpha'V^\mu\gamma^m D_m\epsilon,\nonumber\\
\delta\lambda^A&=&F^A\epsilon,\quad\delta F^A=i\epsilon\gamma^mD_m\lambda^A.
\label{susyrules}
\eea
At this stage, the auxiliary fields $F^A$ are usually integrated out;
we indeed set $F^{A'}=0$, but keep $F\equiv F^{32}$, as it plays a rather
notrivial role in the dynamics of the tachyon condensation.  Indeed, the
tachyon vertex operator with spacetime momentum $p_\mu$ is related to the lone
fermion supermultiplet, and given (in picture~0) by
\be
\CV=\left(F+\lambda\,p_\mu\psi^\mu\right)\exp(ip_\mu X^\mu).
\ee

We now turn on a condensate of the tachyon in the form of worldsheet
superpotential
\be
\label{supp}
S_W=-\frac{\mu}{\pi\alpha'}\int d^2\sigma\left\{F\CT(X)-i\lambda
\psi^\mu\p_\mu\CT(X)\right\},
\ee
and choose $\CT(X)$ such that (\ref{supp}) is an exactly marginal
deformation in conformal gauge.  This requires
\be
\CT(X)=\exp(k_+X^+)\quad {\rm and}\quad -k^2+2V\cdot k=\frac{2}{\alpha'}.
\ee
The real constant $k_+>0$ is thus related to the dilaton gradient via
\be
V_-k_+=-\frac{1}{2\alpha'}.
\ee
These conditions arise at the first and second order of conformal
perturbation theory in the dimensionless $\mu$.

As the next step, we need to choose a gauge.  In critical string
theories with worldsheet supersymmetry, superconformal gauge -- in
which $e_m{}^a=\delta_m{}^a$ and the gravitino $\chi_m$ is set to zero -- is
virtually always chosen.  The residual superconformal
symmetry of this gauge is sometimes further fixed by going additionally
to lightcone gauge; however, this extra gauge fixing is supplemental to
the conformal gauge and not an alternative to it.

To motivate a meaningful alternative gauge, consider the transformation
properties of the lone fermion $\lambda$ in (\ref{susyrules}).  In the
presence of the tachyon condensate, $F$ develops a vacuum expectation value,
and can be integrated out using its algebraic equation of motion,
\be
\label{cond}
F=2\mu\exp(k_+X^+).
\ee
In the rest of this letter, we use $F$ as a shorthand for the
composite operator (\ref{cond}).

In the presence of the condensate, the supersymmetry transformation
$\delta\lambda=F\epsilon$ indicates that $\lambda$ may play the role of
a goldstino for spontaneously broken worldsheet supersymmetry.  Since
this symmetry is local, this in turn suggests that a super-Higgs mechanism
takes place, in which the gravitino develops a propagating degree of freedom.
As a result, it may be wise not to set the gravitino to zero by gauge choice,
and an alternative to the superconformal gauge is required.

In higher dimensions, the Higgs mechanism is usually associated with
the gauge field becoming massive.  In our case, no mass scale should
be generated if the worldsheet theory is to stay conformal. In our
view, the more general signature of the Higgs mechanism is that the
gauge field acquires more physical polarizations.  In four spacetime
dimensions, this happens if a massless gauge field develops a mass.
The worldsheet situation is different. In the absence of tachyon
condensation, the worldsheet gravitino imposes a constraint, setting
the supercurrent to zero on physical states; it effectively carries
``minus one'' polarization. When the tachyon condensate develops,
the worldsheet gravitino will acquire a propagating polarization:
The net number of polarizations increases, but -- unlike in four
dimensions -- no mass is acquired.

Our alternative gauge for worldsheet supersymmetry still involves fixing
worldsheet diffeomorphisms by the conventional conformal gauge,
\be
\label{cg}
e_m{}^a=\delta_m{}^a.
\ee
This is consistent with our expectation of worldsheet conformal invariance.
From now on, our analysis continues in this gauge; and we use worldsheet
lightcone coordinates $\sigma^\pm=\tau\pm\sigma$.  In these coordinates, the
only nonzero components of the fermi fields are $\chi_{++}$, $\psi_-^\mu$,
and $\lambda_+^A$.

Since the gravitino is expected to gain a physical polarization at
the expense of the goldstino $\lambda$, the first instinct for a
better gauge might be to set $\lambda=0$.  The equation of motion
that follows from varying $\lambda$ would then be imposed as a
constraint on physical states.  In our case, this would imply
\be
\mu\psi_-^+\exp(k_+X^+)=0,
\ee
which can be solved by setting $\psi_-^+=0$.  This would in turn require that
the equation of motion obtained from varying $\psi_-^+$ becomes a new
constraint.

The benefit of such an alternative gauge choice is that the gravitino is not
artificially set to zero, and is free to develop its own dynamics.
Once we set $\psi_-^+=0$ in the full action, as tenatively suggested by our
first look at this alternative gauge above, the gravitino couples to the rest
of the system in a relatively simple way, via the terms
\be
\label{terms}
\chi_{++}\psi_-^-\p_-X^++2\alpha'V_-\chi_{++}\p_-\psi_-^-
-2\chi_{++}\psi^i\p_-X^i
\ee
in the action.  This is further simplified by rescaling
\be
\label{resgrav}
\tilde\chi_{++}=F\chi_{++},\qquad\tilde\psi_-^-=\frac{\psi_-^-}{F}.
\ee
This rescaling changes the conformal weights of these fields from $(1,-1/2)$
and $(0,1/2)$ to $(3/2,0)$ and $(-1/2,0)$ respectively, and transforms
(\ref{terms}) to
\be
2\alpha'V_-\tilde\chi_{++}\p_-\tilde\psi_-^--\frac{2}{F}
\tilde\chi_{++}\psi^i\p_-X^i.
\ee
Thus, $\tilde\chi_{++}$ and $\tilde\psi_-^-$ appear to form a left-moving
first-order system of spin $3/2$ and $-1/2$ and central charge $c_L=-11$,
whose coupling to the transverse degrees of freedom $X^i, \psi^i$ becomes
arbitrarily weak at late $X^+$.  This left-moving first-order system is
our dynamical gravitino sector!

In the precise implementation of this alternative gauge, a slight modification
of the process outlined above is useful.  Thus, as our gauge choice, we choose
to supplement (\ref{cg}) by setting
\be
\label{pixiegauge}
\psi_-^+=0,
\ee
instead of setting $\lambda_+=0$ (or $\chi_{++}=0$ as in superconformal
gauge).  As we will see, this accomplishes all the expected benefits outlined
above, and leads to other simplifications.

The Faddeev-Popov superdeterminant $\Delta_{\rm FP}$ for our gauge choice
factorizes, $\Delta_{\rm FP}=J_{bc}/J_{\psi_-^+\epsilon}$.  Here $J_{bc}$ is
the determinant associated with the conformal gauge for worldsheet
diffeomorphisms, traditionally realized by the path integral over a $bc$
ghost system of spin 2 and central charge $c_L=c_R=-26$.
$J_{\psi_-^+\epsilon}$ is the determinant of the operator associated with the
change of variables in the fermionic sector, from $\psi_-^+$ to $\epsilon$,
as determined from the supersymmetry variation of the gauge-fixing condition,
\bea
\label{trans}
\delta\psi^+&=&\gamma^m\p_mX^+\epsilon-2\alpha'V_-\gamma^mD_m\epsilon
\nonumber\\
&=&\frac{4\alpha'V_-}{F}D_-(F\epsilon).
\eea
Hence, $J_{\psi_-^+\epsilon}$ is the determinant of
\be
\label{oper}
D_F\equiv \frac{1}{F}D_-F,
\ee
acting on fields of spin $j=1/2$.  When properly regularized and
renormalized, this determinant will depend not only on the Liouville
field $\phi$ but also on $X^+$ due to the explicit appearance of
$F$ in (\ref{trans}).  This determinant can be explicitly evaluated
using formulas found for example in \cite{kallosh} (see \cite{pixie}
for details).  It is useful to consider the slight generalization of 
$D_F$ given by (\ref{oper}) as acting on fields of spin $j$;  the 
determinant of the corresponding Laplacian is
\bea
\label{FPdet}
&&\frac{1}{2}\log\,\det(D^\dagger_F D_F)\nonumber\\
&&{}=-\frac{1}{48\pi}\int d^2\sigma\,\hat e\left\{[3(2j-1)^2-1]
\hat h^{mn}\p_m\phi\p_n\phi\right.\nonumber\\
&&\qquad\qquad +12(2j-1)k_+\hat h^{mn}\p_m\phi\p_n X^+\nonumber\\
&&\quad\qquad\qquad\qquad\left.{}+12k_+^2\hat h^{mn}\p_mX^+\p_nX^+\right\},
\eea
where $h_{mn}=e^{2\phi}\hat h_{mn}$, and $\hat h_{mn}$ is a fiducial metric.

At $j=1/2$, the first term can be realized by introducing a bosonic
ghost-antighost pair, $\tilde\beta$, $\tilde\gamma$.  These are
standard left-moving first-order fields of conformal weight
$(1/2,0)$.  The second term in (\ref{FPdet}) vanishes for $j=1/2$,
and the third, $X^+$ dependent term is a one-loop contribution to
the action of the $X^\pm$ system.  Note that the ghosts $\tilde\beta$,
$\tilde\gamma$ associated with our gauge choice are purely
{\it left-moving\/}, and contribute $c_L=-1$ to the {\it bosonic\/} sector
on the worldsheet.  This is to be contrasted with conventional superconformal
ghosts in superconformal gauge, which are {\it right-moving\/} and contribute
$c_R=11$ to the supersymmetric sector instead.

In our alternative gauge, the worldsheet theory thus looks very different
than in superconformal gauge.  We have gained a left-moving propagating
massless gravitino of spin 3/2, which (together with its conjugate partner
$\tilde\psi_-^-$ of spin $-1/2$) contributes central charge $c_L=-11$.
We have also gained left-moving bosonic ghosts of spin 1/2, contributing an
additional $-1$ to $c_L$; altogether we have lost twelve units of left-moving
central charge. Similarly, in the right-moving sector we have lost
the $\psi_-^\pm$ pair of fermions, worth $c_R=1$: $\psi_-^+$ is zero
by our gauge choice, and $\psi_-^-$ has turned into a left-moving
field.  In addition, no conventional right-moving superconformal
ghosts $\beta, \gamma$ of central charge $c_R=11$ arise, leading to
the same apparent total discrepancy of 12 units of central charge as
in the left-moving sector.

Does this mean that the theory lost its conformal invariance?  The answer
is no: A careful analysis of the changes of variables ({\it i.e.}, the gauge
fixing and the $X^+$ dependent rescaling of fields) reveals one-loop
corrections due to the measure factors that precisely shift the dilaton from
its originally null direction, making up for the missing twelve units of
central charge and restoring quantum conformal invariance.  The one-loop
correction from the Faddeev-Popov determinant to the effective action
has been calculated above, and is given by
\be
\label{xx}
-\frac{1}{4\pi}\int d^2\sigma\,\hat e\,k_+^2\hat h^{mn}\p_mX^+\p_nX^+.
\ee
Another similar contribution comes from the rescaling of the fields in the
gravitino sector (\ref{resgrav}). This change has led to a canonical kinetic
term in the classical action, but it did not produce the canonical measure
for such first-order fields:  The original measure of the $\chi_{++}$ and
$\psi_-^+$ fields maps to a measure with an explicit $X^+$ dependence for the
rescaled fields.  We wish to replace this $X^+$ dependent measure by the
canonical one. This generates a conversion factor, which can be evaluated
by comparing the Gaussian part of the path integral in the original
$\chi_{++}, \psi_-^-$ variables, to the path integral in the new variables
$\tilde\chi_{++}, \tilde\psi_-^-$ {\it with respect to the canonical measure
independent of $X^+$}.  The former is obtained from (\ref{FPdet}) for
$j=3/2$, and is given by
\bea
\label{gravdet}
&&\log\int\CD\chi_{++}\,\CD\psi_-^-\,e^{-S_2(\chi_{++},\psi_-^+,X^+)}
\nonumber\\
&&\qquad{}=-\frac{1}{48\pi}\int d^2\sigma\,\hat e\left\{11\hat h^{mn}\p_m\phi
\p_n\phi\right.\\
&&\left.{}+24k_+\hat h^{mn}\p_m\phi\p_n X^++12k_+^2\hat h^{mn}\p_mX^+
\p_nX^+\right\},\nonumber
\eea
where $S_2$ is the part of the action bilinear in $\chi_{++}$ and
$\psi_-^-$. The integral over $\tilde\chi_{++}$ and $\tilde\psi_-^-$
(with the measure independent of $X^+$) reproduces correctly the
Weyl anomaly represented by the first, Liouville-dependent term; this
confirms that these fields represent a left-moving first-order system with
$c_L=-11$.  The remaining two terms in (\ref{gravdet}) are the conversion
factor due to the replacement of the measure in the gravitino sector by the
canonical one.  The third term cancels the $X^+$ dependent
contribution (\ref{xx}) of the Faddeev-Popov determinant.  The second
one then shifts the linear dilaton term in the action (which we again
write in Minkowski signature), by
\be
\label{loopdil}
\Delta S=-\frac{1}{4\pi}\int d^2\sigma\, e\,k_+X^+R.
\ee
Here we have integrated the $\p\phi\p X^+$ term in (\ref{gravdet})
by parts, and used the expression $e R=-2\hat e\hat\nabla^2\phi$ for
the scalar curvature in conformal gauge.   The one-loop correction
(\ref{loopdil}) makes the dilaton spacelike, and provides the missing
$c=12$ units of the central charge.  The worldsheet theory is conformal at
the quantum level.

The proper implementation of the BRST quantization involves the BRST partner
$B$ of the antighost $\tilde\beta$.  The  integration over $B$ produces a
delta function which imposes the gauge fixing condition.  Once $B$ is
integrated out, it needs to be eliminated from the BRST transformation rule
for $\tilde\beta$.  The new BRST variation of $\tilde\beta$ gives essentially
the equation of motion resulting from the variation of $\psi_-^+$, and must
vanish on physical states.  This constraint has a clear intuitive
meaning:  It takes the form $\sim\mu\lambda_+\exp(k_+X^+)=\ldots$, where the
$\lambda_+$ term comes from the variation of the superpotential, and
``$\ldots$'' are the contributions from all the other terms, none of which
depends on $\lambda_+$.  This simply means that the goldstino $\lambda_+$ is
not a separate dynamical field, but can be solved for in terms of the
other fields, including the now-dynamical gravitino.  This is how the
super-Higgs mechanism is implemented in this gauge.

Note also that our gauge fixing condition leaves no residual supersymmetry
at finite $\mu$.  Thus, in our alternative gauge, the gauge-fixed theory is
not superconformal, even though manifest conformal invariance is maintained.

The picture of the heterotic $E_8$ model in our alternative gauge can be
easily extended to other models with worldsheet supersymmetry.  For example,
in Type 0 string theory in the presence of null linear dilaton $\Phi=V_-X^-$
and an exponential tachyon condensate along $X^+$ (studied extensively in
\cite{hs3}), we can choose a gauge by setting both $\psi_-^+$ and $\psi_+^+$
(the left-moving and right-moving superpartners of $X^+$) to zero.  Compared
to the heterotic model, we expect the entire gauge-fixing procedure to be
doubled, resulting in spin 1/2 bosonic ghosts $\tilde\beta$, $\tilde\gamma$
and a propagating gravitino sector $\tilde\chi$, $\tilde\psi^-$ in the
left-moving as well as right-moving sector.  Before the one-loop effects of
the Faddeev-Popov determinant and of the rescaling of the gravitino sector are
incorporated, the apparent discrepancy in the central charge is $-24$ for each
movers.  The one-loop determinants then shift the dilaton by twice the
heterotic amount, adjusting the central charge by $24$ and restoring exact
conformal invariance at the quantum level.  In the process, the worldsheet
gravitino has again been liberated, leading to a free fermion system of
$c=-11$.  This elucidates some of the results of \cite{hs3} (as well as the
related construction of \cite{bv}), in which a fermionic $c=-11$ system was
found in superconformal gauge.  Our gauge shows that this system shoud be
interpreted as the dynamical worldsheet gravitino.

The two examples of tachyon condensation considered above suggest that
alternative worldsheet gauge choices may illuminate a more general
class of string backgrounds out of equilibrium.  Yet, the physics of
observables in any given background should be independent of the
choice of gauge, suggesting a new type of worldsheet duality in
string theory:   A duality between different gauge choices of the
same worldsheet theory.  This clearly extends the class of
perturbative string dualities encountered so far: Instead of
representing one CFT in two dual ways, we can now look at one string
background in two gauges, in which they are not even described by
the same CFT.  This feature appears to be conceptually new, and we
hope that it will be useful in understanding string backgrounds with
substantial time dependence, beyond the example of heterotic tachyon
condensation studied here.

This work has been supported by NSF Grants PHY-0244900 and PHY-0555662, DOE
Grant DE-AC03-76SF00098, and the Berkeley Center for Theoretical Physics.
One of us (C.A.K.) has also been supported by an NSF Graduate Fellowship.
\bibliography{8}
\end{document}